\documentclass[12pt,preprint]{aastex}

\usepackage{amsmath}
\usepackage{graphicx}
\usepackage{amssymb}
\usepackage{epstopdf}
\shorttitle{Intrinsic brightness of SDSS objects}
\shortauthors{J. B. Miller and T. E. Miller}

\begin{document}

\title{Intrinsic brightness of SDSS objects\\ is similar at all redshifts in de Sitter space}

\author{John B. Miller}
\affil{Western Michigan University, Kalamazoo, MI 49008, USA}
\email{john.b.miller@wmich.edu}

\author{Thomas E. Miller}
\affil{108 Titleist Drive, Bluefield, VA 24605, USA}
\email{tommy81@alumni.princeton.edu}

\begin{abstract}
The redshift-luminosity distributions for well-defined galaxies and quasars in the Sloan Digital Sky Survey (SDSS) are compared for the two redshift-distance relations of a Hubble redshift and a de Sitter redshift.  Assuming a Hubble redshift, SDSS data can be interpreted as luminosity evolution following the Big Bang.  In contrast, given a de Sitter redshift, the intrinsic brightness of objects at all redshifts is roughly the same.  In a de Sitter universe, 95 per cent of SDSS galaxies and quasars fall into a magnitude range of only 2.8, and 99.7 per cent are within 5.4 mag. The comparable Hubble luminosity ranges are much larger:  95 per cent within 6.9, and 99.7 per cent within 11.5 mag.  De Sitter space is now widely discussed, but the de Sitter redshift is hardly mentioned.
\end{abstract}

\keywords{astronomical data bases: surveys -- galaxies: distances and redshifts -- cosmology: observations -- cosmology: theory -- physical data and processes: relativity -- general: history and philosophy of astronomy}

\section{Introduction}

Of late there has been an increase in the number of papers dealing with de Sitter space.  The Maldacena conjecture or anti-de-Sitter space/conformal field theory (AdS/CFT) correspondence and the holographic principle have attracted interest \citep{jm,ew,mt,gkp}, and there are earlier examples \citep{ct,fd,ap}.  However, there has been little discussion of the feature of de Sitter theory that was of most interest during its heyday in the 1920Õs:  the de Sitter redshift.

Part of the reason for this is that the Hubble redshift has been extensively studied and is widely accepted. Supernovae studies have recently provided data consistent with the Hubble redshift.
%
%
The de Sitter redshift was abandoned by its own author \citep{ds2} in favor of the time-dependent solution of \citet{gl}.
%
%
With few exceptions  \citep{gsh,mm}, the de Sitter redshift has not been seriously reconsidered.
%
%

However, it may be possible that the world is actually, not merely asymptotically, of a de Sitter nature.  We reconsider the de Sitter redshift in its original astronomical context.

\section{De Sitter Geometry}
Much of the following is adapted from the original paper by \citet{ds1}.  
We retain de Sitter's original notation as much as possible, even though this notation
may somtimes conflict with modern conventions. We adopt these definitions:

\begin{align*}
  \theta ,\psi ,\omega ,\zeta ,\chi  &\equiv \textrm{angular coordinates}, \\
  x,y,z,u &\equiv {\textrm{Euclidean coordinates}}, \\
  t,\tilde t &\equiv {\textrm{time coordinates}}, \\
  r_1  &\equiv {\textrm{radial pseudo-Euclidean coordinate}}, \\
  r &\equiv {\textrm{radial elliptical coordinate}}, \\
  h &\equiv {\textrm{radial hyperbolic coordinate}}, \\
 {\textrm{R}} &\equiv {\textrm{radius of curvature}}, \\
  \lambda  &\equiv {\textrm{cosmological constant}}, \\
  \rho  &\equiv {\textrm{mass density}}, \\
  \kappa  &\equiv {\textrm{gravitational constant}}.
\end{align*}
In two-dimensional, negatively-curved spacetime, the line element is
\begin{equation*}
\textrm{d}s^2  =  - {\textrm{R}}^{2} \left( {\textrm{d}\psi ^2 { + }\sin ^2 \psi \textrm{d}\theta ^2 } \right).
\end{equation*}
The three-dimensional version is
\begin{equation*}
\textrm{d}s^2  =  - {\textrm{R}}^2 \left[ {\textrm{d}\zeta ^2  + \sin ^2 \zeta \left({\textrm{d}\psi ^2  + \sin ^2 \psi \textrm{d}\theta ^2 } \right)} \right].
\end{equation*}
The four-dimensional version is
\begin{equation}      \label{m1}
\textrm{d}s^2  =  - {\textrm{R}}^2 \left\{ {\textrm{d}\omega ^2  + \sin ^2 \omega \left[ {\textrm{d}\zeta ^2  + \sin ^2 \zeta \left({\textrm{d}\psi ^2  + \sin ^2 \psi \textrm{d}\theta ^2 } \right)} \right]} \right\}.
\end{equation}
This satisfies EinsteinÕs gravitational field equations
\begin{equation*}
G_{\mu \nu }  - \lambda g_{\mu \nu }  =  - \kappa T_{\mu \nu }  + \tfrac{1}{2}g_{\mu \nu } T
\end{equation*}
with
\begin{equation*}
{\textrm{R}}^2  = \frac{3}
{\lambda } = \frac{{3c^2 }}
{{8{\pi }\kappa \rho }}.
\end{equation*}
In order to avoid imaginary angles, we can substitute
\begin{align*}
  \omega  &= {\textrm{i}}\omega ',  \\
  \zeta  &= {\textrm{i}}\zeta '. 
\end{align*}
Then the line-element becomes
%
%
%
%
\begin{equation}      \label{m2}
\textrm{d}s^2  =  {\textrm{R}}^2 \left\{ {\textrm{d}\omega '^2  - \sinh ^2 \omega ' \left[ {\textrm{d}\zeta '^2  + \sinh ^2 \zeta '\left({\textrm{d}\psi ^2  + \sin ^2 \psi \textrm{d}\theta ^2 } \right)} \right]} \right\}.
\end{equation}
Transforming to pseudo-Euclidean coordinates 
\begin{align*}
  r_1  &= \textrm{R}\sinh \omega '\sinh \zeta ',&&\tilde t= \textrm{R}\sinh \omega '\cosh \zeta ', \\
  x &= r_1 \sin \psi \sin \theta , \\
  y &= r_1 \sin \psi \cos \theta ,&&u =  \textrm{R}\cosh \omega ', \\
  z &= r_1 \cos \psi ,
\end{align*}
we have
\begin{equation}      \label{m3}
{\textrm{d}}s^2  =  - {\textrm{d}}x^2  - {\textrm{d}}y^2  - {\textrm{d}}z^2  + {\textrm{d}}\tilde t^2  - {\textrm{d}}u^2 
\end{equation}
and
\begin{equation*}
\textrm{R}^2  - x^2  - y^2  - z^2  + \tilde t^2  - u^2  = 0.
\end{equation*}
Then by the transformation
\begin{align*}
  z_1  &= {\textrm{i}}x, \\
  z_2  &= {\textrm{i}}y, \\
  z_3  &= {\textrm{i}}z, \\
  z_4  &= \tilde t, \\
  z_5  &= {\textrm{i}}u ,
\end{align*} 
we obtain the result
\begin{equation}      \label{m4}
{\textrm{d}}s^2  = {\textrm{d}}z_1^2  + {\textrm{d}}z_2^2  + {\textrm{d}}z_3^2  + {\textrm{d}}z_4^2  + {\textrm{d}}z_5^2  ,   
\end{equation}
where
\begin{equation}      \label{z5}
z_1^2  + z_2^2  + z_3^2  + z_4^2  + z_5^2  = \left( {{\text{i}}\textrm{R}} \right)^2 .
\end{equation}
The de Sitter metrics given by equations~\eqref{m1}, \eqref{m2}, \eqref{m3}, and \eqref{m4} 
demonstrate that in de Sitter spacetime, there is no preferred origin in either space or time, but with these metrics, time and space are mingled so that conventional rulers and clocks are not useful.
\begin{quotation}
[Equation~\eqref{z5} is] the equation which determines that four-dimensional surface in the five-dimensional manifold that corresponds to space-time.  In accordance with this result we can regard the geometry of the de Sitter universe as that holding on the surface of a sphere embedded in five-dimensional Euclidean space.  And, as in the case of the Einstein universe, we gain an added intuitional appreciation of the homogeneity of the de Sitter model.  It may be emphasized, nevertheless, that the formal simplicity in the expression for the line element given by [equation~\eqref{z5}] is achieved at the expense of losing track of the physical distinction between space-like intervals which are to be measured in principle by the use of metre sticks and time-like intervals which are measurable with the help of clocks. \citep{rct}
\end{quotation}
Static de Sitter metrics for which the $g_{\mu \nu }$ are independent of the time coordinate are of more physical interest.
\begin{quotation}
In both systems A [Einstein] and B [de Sitter] it is always possible, at every point of the four-dimensional time-space, to find systems of reference in which the $g_{\mu \nu }$ depend only on one space-variable (the ``radius-vector''), and not on the ``time.'' In the system A the ``time'' of these systems of reference is the same always and everywhere, in B it is not. In B there is no universal time; there is no essential difference between the ``time'' and the other three coordinates.  None of them has any real physical meaning. \citep{ds1}
\end{quotation}
To obtain a more traditional de Sitter line element, we transform coordinates
%
%
%
\begin{align*}
  \tilde t &= \sqrt {\textrm{R}^2  - r_1^2 } \sinh \left( {\frac{t}
{\textrm{R}}} \right), \\
  u &= \sqrt {\textrm{R}^2  - r_1^2 } \cosh \left( {\frac{t}
{\textrm{R}}} \right), 
\end{align*}
yielding the well-known pseudo-Euclidean de Sitter metric corresponding to the inside of a sphere $r_1  \leq \textrm{R}$ in Euclidean space,
\begin{equation}      \label{euclidean}
\textrm{d}s^2  =  - \left( {1 - \frac{{r_1^2 }}
{{\textrm{R}^2 }}} \right)^{ - 1} {\textrm{d}}r_1^2  -  r_1^2 \left[ {{\textrm{d}}\psi ^2  + \sin ^2 \psi {\textrm{d}}\theta ^2 } \right] + \left( {1 - \frac{{r_1^2 }}
{{\textrm{R}^2 }}} \right){\textrm{d}}t^2 
\end{equation}
(choosing units so that $c = 1$).
This is one of the most common forms of the static (time-independent) de Sitter metric. It is less symmetric, but in some ways more intuitive than equation~\eqref{m1} or equation~\eqref{m4}.  Time and space have been disentangled, imaginary coordinates have been rendered real, and the metric is time-independent:  the $g_{\mu \nu }$ are a function of only the radial coordinate and not the time coordinate.  The de Sitter metric given by equation~\eqref{euclidean} is unique in that the surface area increases with the square of the radial coordinate $r_1$ so that luminosity distance $D_\textrm{L}$ is given simply by
\begin{equation}      \label{dl}
D_\textrm{L} = r_1.
\end{equation}
By the transformation
\begin{equation*}
r_1  = \textrm{R}\sin \frac{r}{\textrm{R}},
\end{equation*}
we obtain the metric of elliptical space
\begin{equation*}      \label{elliptical}
{\textrm{d}}s^2  =  - {\textrm{d}}r^2  - \textrm{R}^2 \sin ^2 \frac{r}
{\textrm{R}}\left[ {{\textrm{d}}\psi ^2  + \sin ^2 \psi {\textrm{d}}\theta ^2 } \right] + \cos ^2 \frac{r}
{\textrm{R}}{\textrm{d}}t^2 ,
\end{equation*}
written more clearly by substituting an angular coordinate
\begin{equation*}      \label{r1}
r = \textrm{R}\chi ,
\end{equation*}
so that
\begin{equation}      \label{chi}
{\textrm{d}}s^2  =  - \textrm{R}^2 \ {\textrm{d}}\chi ^2  - \textrm{R}^2 \sin ^2 \chi \left[ {{\textrm{d}}\psi ^2  + \sin ^2 \psi {\textrm{d}}\theta ^2 } \right] + \cos ^2 \chi {\textrm{d}}t^2.
\end{equation}
This elliptical de Sitter metric suggests an interesting topology for de Sitter space, whereby antipodal points are identified, and the whole of space is mapped on to one-half of a one-sided hypersphere.  This topological mapping can be intuitively represented in the two-dimensional case as the surface of a hemisphere in three-dimensional space with cross-connectivity.

For pseudo-Euclidean or elliptical de Sitter coordinates, the velocity of light is not constant. For example, given equation \eqref{chi} the radial velocity of light $v = \cos{\chi}$.
If we introduce a new variable $h$ by the condition

\begin{equation*}
\frac{{{\textrm{d}}r}}{{{\textrm{d}}h}} = \cos \chi,
\end{equation*}
of which the integral is
\begin{equation*}
\sinh \frac{h}{\textrm{R}} = \tan \frac{r}{\textrm{R}},
\end{equation*}
the velocity of light will be constant in all directions \citep{ds1}. The line element becomes
\begin{equation}      \label{hyperbolic}
{\textrm{d}}s^2  = \frac{{ - {\textrm{d}}h^2  - \textrm{R}^2 \sinh ^2 \frac{h}
{\textrm{R}}\left[ {{\textrm{d}}\psi ^2  + \sin ^2 \psi {\textrm{d}}\theta ^2 } \right] + {\textrm{d}}t^2 }}
{{\cosh ^2 \frac{h}
{\textrm{R}}}}
\end{equation}
The three-dimensional space of this system of reference is the space with constant negative curvature:  the space of \citet{nil} and \citet{bo}.  This coordinate system is most natural if one takes the essence of relativity to be the inability of an observer to measure a difference in the speed of light.

Unfortunately, for the case of negative curvature there is no intuitive, isometric mapping of a complete, negatively-curved, two-dimensional surface in three-dimensional Euclidean space \citep{dh}. Mapping a two-dimensional space with negative curvature on to a surface in three-dimensional Euclidean space allows some features of the geometry to become intuitive, but only at the expense of other features that become distorted.

The various de Sitter metrics are useful in different ways.  The metric given by equation~\eqref{m1} demonstrates that all points in the geometry are equivalent. The pseudo-Euclidean metric given by equation \eqref{euclidean} is unique in that the surface area increases with the square of the radial coordinate $r_1$. The elliptical metric given by equation \eqref{chi} has an interesting topology.  And the hyperbolic metric given by equation \eqref{hyperbolic} has the unique property that the speed of light is the same at all places in all directions.

Typical de Sitter $g_{\mu \nu }$ are given in Table 1. Note that for static de Sitter coordinates,
\begin{equation*}
g_{22}  = \textrm{R}^2 \left( {g_{44}  - 1} \right),
\end{equation*}
and that
\begin{equation*}
g_{44}  = \left( {z + 1} \right)^{ - 2}.
\end{equation*}
\renewcommand{\arraystretch}{1.5}

\begin{table}

\caption{De Sitter geometry.}

\begin{tabular}{l c c c}

\hline 
Geometry & Pseudo-Euclidean & Elliptical & Hyperbolic  \\[1ex]

\hline 

Metric  &  Eq. \eqref{euclidean}&  Eq. \eqref{chi}&  Eq. \eqref{hyperbolic}  \\ [1ex]

Coordinates  &   $r_1 ,\psi ,\theta ,t$  &   $\chi ,\psi ,\theta ,t$  &   $h ,\psi ,\theta ,t$   \\ [1ex]
$g_{11}$  &  $ - \left( {1 - \frac{{r_1^2 }} {{\textrm{R}^2 }}} \right)^{ - 1} $  &
                        $- \textrm{R}^2$   &
                        $-$ sech$^2 {\frac {h}{\textrm{R}}}$       \\ [2ex]
                        
$g_{22}$   &  $ - r_1^2 $  &  $ - \textrm{R}^2 \sin ^2 \chi $ 
                    &  $ - \textrm{R}^2 \tanh ^2 \frac{h} {\textrm{R}} $        \\ [2ex]

$g_{44}$  &  $  \left( {1 - \frac{{r_1^2 }} {{\textrm{R}^2 }}} \right) $
                   &  $\cos ^2 \chi $   &
                        sech$^2 {\frac {h}{\textrm{R}}}$      \\ [2ex]

$\frac{{g_{22} }} {{\textrm{R}^2 \left( {g_{44}  - 1} \right)}}$  & 1 & 1 & 1  \\ [1ex]

\hline

\end{tabular}
\end{table}
The ratio of the observed flux $F_\textrm{A}$ from an object at a distance $\textrm{A}$ with redshift $z_\textrm{A}$, and the observed flux $F_\textrm{B}$ from a similar object at a distance $\textrm{B}$ with redshift $z_\textrm{B}$, is given by
\begin{equation}       \label{flux}
\frac{{F_\textrm{A} }}
{{F_\textrm{B} }} = \frac{{g_{22}^\textrm{B} }}
{{g_{22}^\textrm{A} }} = \frac{{1 - g_{44}^\textrm{B} }}
{{1 - g_{44}^\textrm{A} }} = \frac{{1 - \left( {z_\textrm{B}  + 1} \right)^{ - 2} }}
{{1 - \left( {z_\textrm{A}  + 1} \right)^{ - 2} }},
\end{equation}
where $g_{\mu \nu }^\textrm{A} $ and $g_{\mu \nu }^\textrm{B} $ are the values of the $g_{\mu \nu }$ at the radial coordinate distances $\textrm{A}$ and $\textrm{B}$ respectively.  Apparent magnitude is defined as
\begin{equation}      
m_\textrm{A}  - m_\textrm{B}  =  - 2.5\log \left( {\frac{{F_\textrm{A} }}
{{F_\textrm{B} }}} \right)
\end{equation}
where $m_\textrm{A}$ and $m_\textrm{B}$ are the apparent magnitudes of obects at distances $\textrm{A}$ and $\textrm{B}$.
In accord with the tensor character of equation~\eqref{flux}, the relationship between magnitude and redshift does not depend on the choice of hyperbolic, elliptical, or pseudo-Euclidean coordinates:  it is independent of the choice of coordinate transformations.  The distance coordinate is effectively eliminated.

No matter which coordinate transformation is chosen, luminosity distance $D_\textrm{L}$ is a function of $g_{22} $,
\begin{equation}    
D_{\textrm{L}}  = \left( { - g_{22} } \right)^{\frac{1}{2}}.
\end{equation}
The coordinate $r_1$ is convenient in that $D_\textrm{L} = r_1$, but the same results will be obtained if the transformations leading to $\chi$ or $h$ are applied consistently. The choice of coordinate systems is largely one of convenience:  the relationship between the two observables ($z$ and $m$) and the derived quantity ($M_\text{de Sitter}$) will not be affected by choice of coordinates.

Several de Sitter metrics have been given above.  There are many other coordinate transformations, but the static de Sitter metric is well represented by equations \eqref{euclidean}, \eqref{chi}, and \eqref{hyperbolic}, corresponding to flat, positively-curved, and negatively-curved space respectively.
\section{Redshift and luminosity}
Large astronomical distances cannot be directly measured but are inferred from redshift according to some redshift-distance law.  Since only apparent magnitudes are actually measured, discussions about distances in astronomy are really discussions about the derived absolute magnitudes.

Absolute magnitude $M$ is related to apparent magnitude $m$ and luminosity distance $D_\textrm{L}$  (with $D_\textrm{L}$ in Mpc) by
\begin{equation}  \label{Mm}
M = m - 5\log D_{\textrm{L}}  - 25
\end{equation}
Although there are many different `nontrivial' redshifts \citep{r}, we focus on two:  the Hubble redshift and the de Sitter redshift.
\subsection{Hubble}
Assuming a Hubble law, the luminosity distance $D_\textrm{L}$ is related to redshift $z$ by
\begin{equation}       \label{DLextended}
D_{\textrm{L}}  = \frac{c}
{{H_0 q_0^2 }}\left[ {1 - q_0  + q_0 z + \left( {q_0  - 1} \right)\left( {2q_0 z + 1} \right)^{\tfrac{1}
{2}} } \right],
\end{equation}
where $H_0$ is the Hubble constant, and $q_0$ is the deceleration constant  \citep{z}. We neglect the K-correction and the correction for interstellar absorption, which are both small, focusing on a general comparison of the Hubble and de Sitter redshifts.  Assuming $q_0 = 1$, equation~\eqref{DLextended} becomes
\begin{equation}     
D_{\textrm{L}}  = \frac{{cz}}
{{H_0 }}.
\end{equation}
Inserting this into equation~\eqref{Mm}, we obtain
\begin{equation}       \label{Mhubble}
M_{{\textrm{Hubble}}}  = m - 5\log (z) + C_{{\textrm{Hubble}}}
\end{equation}
where
\begin{equation}     
C_{\textrm{Hubble}}  = -5 \log{\frac{{c}}
{{H_0 }}} - 25
\end{equation}
with $H_0$ in $\textrm{km}~s^{-1}~\textrm{Mpc}^{-1}$.

Assuming $H_0 = 75$ and $H_0 = 50$, we have respectively, $C_{\textrm{Hubble}}  = -43.0$ and 
$C_{\textrm{Hubble}}~=-43.9$.
A different value of $H_0$ would shift the data by a fixed amount but will not affect the slope.  For example, a change from $H_0 = 75$ to $H_0 = 50$ would make all objects brighter by
\begin{equation*}     
5\log{(50/75)} = -0.88~\textrm{mag}.
\end{equation*}
\subsection{De Sitter}
Assuming photons are emitted at a constant rate in all directions from similar objects at different radial distances, the flux from an object at a given coordinate distance will be inversely proportional to the surface area of the space at that distance.  In Euclidean space, this gives rise to the familiar inverse-square law dimming since surface area increases as the distance squared.  However, in non-Euclidean space, surface area is not proportional to the square of the radius.

Using Table 1, the de Sitter line element may be written  as
\begin{equation}       \label{generalMetric}
\textrm{d}s^2  = g_{11} {\textrm{d}}r^2  + g_{22} \left[ {{\textrm{d}}\psi ^2  + \sin ^2 \psi {\textrm{d}}\theta ^2 } \right] + g_{44} c^2 {\textrm{d}}t^2 
\end{equation}
so that surface area is proportional to $g_{22}$.  
Only the metric of equation~\eqref{euclidean} has inverse-square law dimming,  as noted above [equation~\eqref{dl}].  The de Sitter redshift can be given by
\begin{equation}       \label{g44}
z = \left( {g_{44} } \right)^{ - \tfrac{1}
{2}}  - 1 = \left( {1 - \frac{{r_1^2 }}
{{\textrm{R}^2 }}} \right)^{ - \tfrac{1}
{2}}  - 1,
\end{equation}
or
\begin{equation}       
r_1  = R\left[ {1 - \left( {z + 1} \right)^{ - 2} } \right]^{\tfrac{1}
{2}} ,
\end{equation}
so that
\begin{equation}      \label{dsm}
M_{\textrm{de Sitter}}  = m - 2.5\log [1 - (z + 1)^{ - 2} ] + C_{\textrm{de Sitter}} ,
\end{equation}
with $C_{\textrm{de Sitter}} =  - 5\log{\textrm{R} - 25}$.
For example, assuming $\textrm{R} = 10^{10}$ light-years (or equivalently density $\rho  = 1.8 \times 10^{ - 29}~\textrm{g~cm}^-3$),
one obtains $C_{\textrm{de Sitter}}=-42.4~\textrm{mag}$.

A determination of the actual value of $C_{\textrm{de Sitter}}$ is equivalent to a determination of the value of $\textrm{R}$, akin to a measurement of $H_0$, and beyond the current scope of this work.  The actual value of $C_{\textrm{de Sitter}}$ is not essential to the analysis, since changing $C_{\textrm{de Sitter}}$ will displace all of the data by a fixed amount.  
\section{Observations}
We extracted galaxies and quasars from the Sloan Digital Sky Survey \citep{ab} that have redshift confidence greater than 0.95.  The raw data (apparent magnitude vs. log redshift) are shown in Fig. 1, and the Hubble [equation \eqref{Mhubble}] and de Sitter [equation~\eqref{dsm}] absolute magnitude transformations are shown in Fig.~2 and Fig.~3, respectively.  We have assumed $C_{\textrm{Hubble}} = -43.0$ and $C_{\textrm{de Sitter}} = -42.4~\textrm{mag}$.  We show only the \textit{R} magnitude, but similar results are obtained with any of the \textit{UGRIZ} magnitudes.  Likewise, the results are not sensitive to the choice of redshift confidence.
\begin{figure}
        \center{\includegraphics [width=\linewidth]
         {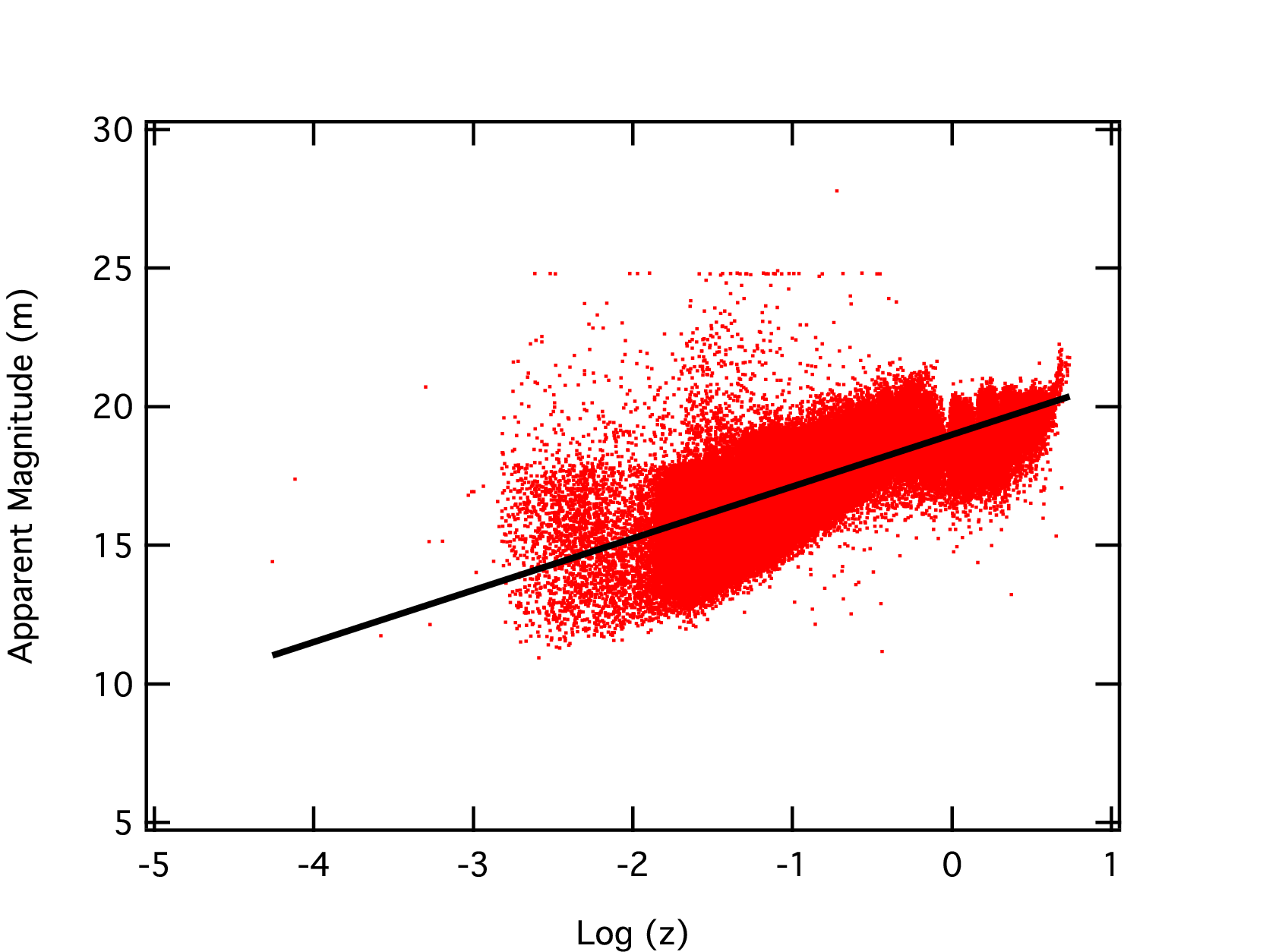}}
        \caption{ \textit{R}-band magnitudes of 786409 SDSS DR7 objects with redshift confidence greater than 0.95 are plotted versus log($z$).
        Of these, 717036 are galaxies, and 69373 are quasars.  The line is a best fit to the data to show the general trend in the dataset.}
\end{figure}

\begin{figure}
        \center{\includegraphics [width=\linewidth]
         {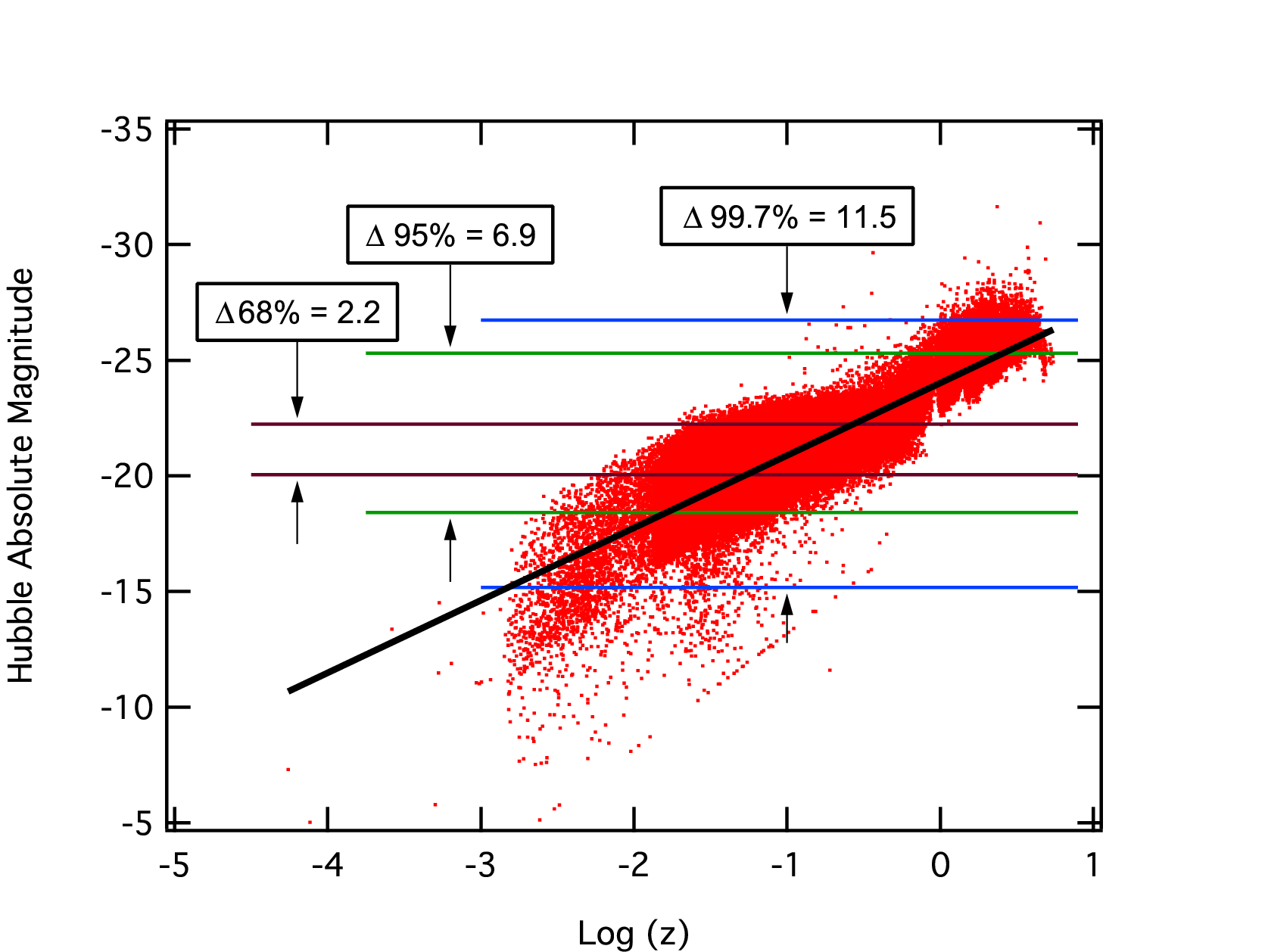}}
        \caption{ Hubble absolute magnitude versus log($z$) for SDSS galaxies and quasars.  There is inreasing intrinsic brightness with inreasing redshift, consistent with luminosity evolution in a Big Bang universe.}
\end{figure}
\begin{figure}
      \center{\includegraphics [width=\linewidth]
         {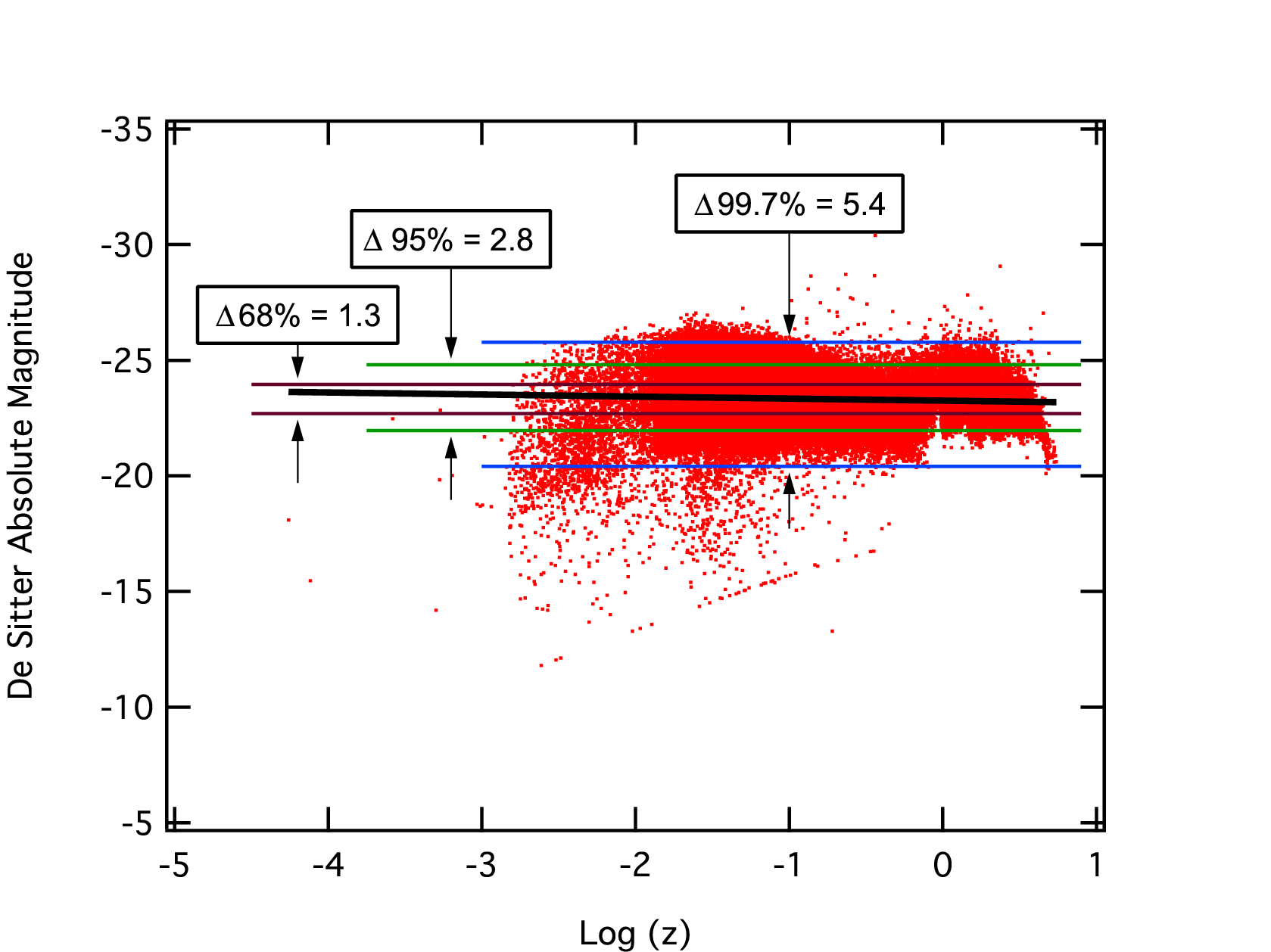}}
        \caption{ De Sitter absolute magnitude versus log($z$) for SDSS galaxies and quasars.  Intrinsic brightness is roughly the same at all redshifts, consistent with a quasi-static de Sitter universe. }
\end{figure}

The upward slope of Fig. 1 shows that, for low-redshift objects, dimness increases with increasing redshift and suggests that redshift is a distance effect, not a local or intrinsic phenomenon.

Assuming a Hubble redshift (Fig. 2), high-redshift objects are intrinsically brighter than low-redshift objects. This trend is currently interpreted as luminosity evolution following the Big Bang.

However, it may be that the data can be interpreted in a different, yet self-consistent way.  Note the narrow range and roughly horizontal configuration of Fig. 3.  Given a de Sitter redshift, the intrinsic brightness of objects at all redshifts is practically the same.

Assuming a de Sitter redshift, 99.7\% of SDSS objects at all redshifts span a range of only 5.4 mag, while assuming a Hubble redshift, the range is more than double at 11.5~mag (Table 2).  This may be a coincidence, but the de Sitter redshift is especially interesting in light of recent work on de Sitter space in allied fields.
\renewcommand{\arraystretch}{1.5}

\begin{table}

\caption{SDSS absolute magnitude range.}

\begin{tabular}{l c c}

\hline 
Percentile & Hubble & De Sitter  \\

\hline 

$\Delta$68\%  &  2.2 &  1.3  \\
$\Delta$95\%  & 6.9  &  2.8  \\
$\Delta$99.7\%  &  11.5 &  5.4  \\

\hline

\end{tabular}
\end{table}
\section{Discussion}

\subsection{Schwarzchild-de Sitter}

The de Sitter metric [equation~\eqref{euclidean}] is the minimal central mass ($M~\to~0$) limiting case of the more general Schwarzchild-de Sitter metric
\begin{equation}       \label{sds}
{\textrm{d}}s^2  =  - \left( {1 - \frac{{2M}}
{{r_1 }} - \frac{{r_1^2 }}
{{R^{2} }}} \right)^{ - 1} {\textrm{d}}r_1 ^2 - 
 r_1 ^2 \left[ {{\textrm{d}}\psi ^2  + \sin ^2 \psi {\textrm{d}}\theta ^2 } \right]   
+  \left( {1 - \frac{{2M}}
{{r_1 }} - \frac{{r_1^2 }}
{{R^{2} }}} \right){\textrm{d}}t^2 . 
\end{equation}
Similarly, the Schwarzchild metric
\begin{equation}       \label{sch}
{\textrm{d}}s^2  =  - \left( {1 - \frac{{2M}}  
{{r_1 }}} \right)^{ - 1} {\textrm{d}}r_1 ^2 - 
 r_1 ^2 \left[ {{\textrm{d}}\psi ^2  + \sin ^2 \psi {\textrm{d}}\theta ^2 } \right] 
+ \left( {1 - \frac{{2M}}
{{r_1 }}} \right){\textrm{d}}t^2 
\end{equation}
is the limiting case of the Schwarzchild-de Sitter metric [equation \eqref{sds}] where $R~\to~\infty$, equivalent to a point-like central mass.

The de Sitter metric [equation \eqref{euclidean}] can be obtained from the Schwarzchild metric [equation \eqref{sch}] by recalling
\begin{equation*}
{\textrm{R}}^2  = \frac{3}
{\lambda } = \frac{{3c^2 }}
{{8{\pi }\kappa \rho }}.
\end{equation*}
and assuming
$M=\tfrac{4}{3}{\pi}r_1^3 \rho $
(choosing units so that $\kappa=c =1$ ). Others have reached similar conclusions \citep{ba}.
\subsection{Interior Schwarzchild}
The de Sitter solution has an interesting relationship to the Schwarzchild interior solution \citep{rct} for a perfect fluid sphere of constant density $\rho$,
\begin{equation}       \label{schint}
{\textrm{d}}s^2  =  - \frac{{{\textrm{d}}r_1 ^2 }}  
{{1 - \frac{{r_1 ^2 }}
{{R^2 }}}} - r_1 ^2 {\textrm{d}}\theta ^2  - r_1 ^2 \sin ^2 \theta {\textrm{d}}\phi ^2  +     
 \left( {A - B\sqrt {1 - \frac{{r_1 ^2 }}
{{R^2 }}} } \right)^2 {\textrm{d}}t^2 
\end{equation}
where $A$ and $B$ are integration constants.  Continuity between the interior  [equation \eqref{schint}]  and exterior  [equation \eqref{sds}]  Schwarzchild-de Sitter metrics at the surface of the sphere, for the limit $M~\to~0$, yields 
$A = 0$ and $B = 1$ .

The interior Schwarzchild solution puts an upper limit on the possible size of a sphere of given density. The size of the de Sitter universe is the limiting size of a perfect fluid sphere with density   equal to the mean mass density of the universe.  The similarity of the interior Schwarzchild metric and the de Sitter metric suggests that the de Sitter universe may be loosely construed as a giant, low-density, inside-out black hole.
\subsection{Finite surface area}
The surface area of the de Sitter universe is finite, even though coordinate distances such as $h$ may be infinite.  In the hyperbolic plane, the area of a maximal triangle has a finite value of $\pi$, while each of the three edges is infinitely long.  Similarly, in de Sitter space, lines may be infinitely long, but surface area is finite.
\subsection{Lightspeed}

Assuming a de Sitter metric, there is no distinction between the three fundamental geometries with respect to the magnitude-redshift relation.  However, direct lightspeed measurement is a local measurement that may bear on the reality of the hyperbolic, Euclidean, and elliptical spaces.  We have found that local lightspeed from redshifted objects is normal \citep{mm2}, and therefore de Sitter space is presumably hyperbolic, or negatively curved.

It might be argued that any measurement of photon velocity will be local, and thereby not cosmologically relevant.  However, such a measurement would be neither more nor less local than the measurement of photon redshift.
\subsection{Galactic rotation}
For a long time, de Sitter space was thought to be elliptical, with positive curvature and radial repulsion.  However, galactic rotation curves and other dark matter phenomena suggest a radial attraction that would be present in de Sitter space with a change in sign of the radius.  The sign of the radial de Sitter acceleration is mathematically somewhat arbitrary.  Negatively curved de Sitter space might help explain the dark matter problem.
\subsection{Supernovae}
Recent work on supernovae has apparently confirmed the Hubble redshift.  One might thereby assume that the de Sitter redshift could not possibly be correct.  However, the supernovae data may have an alternative, self-consistent interpretation.

Current analysis assumes a time dilation factor of $1~+~z$.  It may be interesting to reanalyze the data without the time dilation factor to see whether the supernovae data are consistent with a de Sitter redshift.
\subsection{Black-body radiation}
The discovery of the cosmic microwave background radiation (CMB) is considered a landmark test of the Big Bang model.  However, the CMB may also be interpreted as a Gibbons-Hawking effect extended to the de Sitter solution \citep{gh}.
\section{Conclusion}
Because the Hubble redshift is linear and very well-established at low redshifts, while the de Sitter redshift is quadratic at low redshifts, some may conclude that the de Sitter interpretation of the SDSS data is an aberration.
However, the de Sitter redshift provides an interesting way to interpret the SDSS data and merits more study, especially given the current general interest in de Sitter theory.

\section*{Acknowledgments}

Funding for the SDSS and SDSS-II has been provided by the Alfred P. Sloan Foundation, the Participating Institutions, the National Science Foundation, the U.S. Department of Energy, the National Aeronautics and Space Administration, the Japanese Monbukagakusho, the Max Planck Society, and the Higher Education Funding Council for England. The SDSS Web Site is http://www.sdss.org/.

The SDSS is managed by the Astrophysical Research Consortium for the Participating Institutions. The Participating Institutions are the American Museum of Natural History, Astrophysical Institute Potsdam, University of Basel, University of Cambridge, Case Western Reserve University, University of Chicago, Drexel University, Fermilab, the Institute for Advanced Study, the Japan Participation Group, Johns Hopkins University, the Joint Institute for Nuclear Astrophysics, the Kavli Institute for Particle Astrophysics and Cosmology, the Korean Scientist Group, the Chinese Academy of Sciences (LAMOST), Los Alamos National Laboratory, the Max-Planck-Institute for Astronomy (MPIA), the Max-Planck-Institute for Astrophysics (MPA), New Mexico State University, Ohio State University, University of Pittsburgh, University of Portsmouth, Princeton University, the United States Naval Observatory, and the University of Washington.

\end{document}